\documentclass[pdftex,twocolumn]{article}
\usepackage[utf8]{inputenc}
\usepackage[T1]{fontenc}
\RequirePackage{mathptmx}      
\RequirePackage{flushend}
\RequirePackage[numbers,sort&compress]{natbib}
\usepackage{amsmath}
\usepackage{amssymb}
\usepackage{graphicx}
\usepackage{multicol}
\usepackage{multirow}
\usepackage{caption}
\usepackage{booktabs}
\usepackage[english]{babel}
\title{All-charm tetraquark mass and possible quantum numbers of $X(6900)$}

\author{Morgan Kuchta\thanks{309312@uwr.edu.pl} \\University of Wrocław, Institute of Theoretical Physics, pl. Maxa Borna 9 Wrocław, 50-204, Poland}
\date{}
\begin{document}

	\maketitle
	
	\begin{abstract}
		In this work we propose possible quantum numbers of $X(6900)$ and suggest a model for it internal structure that explains its unusually high mass. We solve the Schr\"o-
		dinger Equation with Mathematica 12, first for charmonium spectrum, then for all-charm tetraquark spectrum which is understood as a pair of two-particle states, mesons or diquark-antidiquark states. The obtained candidates for all-charm te-traquark will be separated into contributors to various resonances and structures that are visible in the experiments then an explanation for the prominence of $X(6900)$ will be proposed.
		
	\end{abstract}
	
	\section{Introduction}
	
	In June 2020 in LHCb experiment found in proton-proton collisions exotic meson originally named $X(6900)$ along with it a broad structure between $6200-6800 \ \text{MeV}$ and a smaller peak around $7200\ \text{MeV}$ were found \cite{LHCb:2020-6900}. Until now, the parity and the charge symmetry, as well as the total angular momentum of any of the states mentioned here were not identified. The mass of $X(6900)$ was surprisingly high, as before its discovery the expected mass of the ground state was placed around $6200\ \text{MeV}$, which means that it should have been found within the region of the broad structure. In 2022, the ATLAS collaboration searched for potential all-charm tetraquark ($cc\bar{c}\bar{c}$) states in two different decay channels \cite{ATLAS2022}. Results of the search also feature a peak around $6900\ \text{MeV}$ and the broad structure visible in results archived in the previous experiment. The experiment differs from the one conducted in 2020 by investigating the $J/\psi+\Psi(2S)$ invariant mass spectrum, where the two significant peaks were found, one supposedly corresponding to the peak around $6900\ \text{MeV}$ found in LHCb, another around $7200\ \text{MeV}$, both of them have statistical significance lesser than $5\sigma$. Other significant results were presented by the CMS collaboration \cite{CMS} that suggests three peaks in the $di-J/\psi$ spectrum. A summary of the results can be found in the Table \ref{DataTable}.

	\begin{table*}[h] \captionof{table}{Summary of estimated masses (in MeV) and widths of four resonances from three different sources. If the source includes several estimations they are included with brief descriptions. More detailed explanations can be found in the sources. Marker "NA" suggests that resonance was either not identified or its mass was not estimated by the model. \label{DataTable}}
		\begin{center}
			\begin{tabular}{llllll}
				\toprule
				& \multicolumn{1}{}{}            & $X(6200)$ & $X(6500)$ & $X(6900)$ & $X(7200)$  \\ \cmidrule{3-6}
				\multicolumn{1}{c}{\multirow{4}{*}{}}  & \multirow{2}{*}{} &     &       &      &      \\[-2.3ex]
				\multicolumn{1}{c}{\multirow{4}{*}{LHCb \cite{LHCb:2020-6900}}}  & \multirow{2}{*}{No interference} &     \multirow{2}{*}{Observed}    &    \multirow{2}{*}{Observed}    & $m=6905   $   MeV   &   \multirow{2}{*}{Observed}        \\ 
				\multicolumn{1}{c}{}                       &                                  &        &        &     $\Gamma=80 $  MeV  &          \\ \cmidrule{2-6}
				\multicolumn{1}{c}{\multirow{4}{*}{}}  & \multirow{2}{*}{} &    &  &      &    \\[-2ex]
				\multicolumn{1}{c}{}                       & \multirow{2}{*}{NRSPS interference}    &    \multirow{2}{*}{Observed}      &   \multirow{2}{*}{Observed}       &      $m= 6886 $  MeV   &  \multirow{2}{*}{Observed}          \\ 
				\multicolumn{1}{c}{\multirow{4}{*}{}}  & \multirow{2}{*}{} &     &       &      &      \\[-2.3ex]
				\multicolumn{1}{c}{}                       &                                  &         &         &     $\Gamma=168 $    MeV &           \\ \midrule
				\multicolumn{1}{c}{\multirow{4}{*}{}}  & \multirow{2}{*}{} &     &       &      &      \\[-2.3ex]
				\multicolumn{1}{l}{\multirow{4}{*}{ATLAS \cite{ATLAS2022}}} & \multirow{2}{*}{Model A}         &    $m=6.22 $ GeV  &     $m=6.62 $ GeV    &   $m=6.87 $ GeV      & $m=7.22 $ GeV          \\ 
				\multicolumn{1}{c}{\multirow{4}{*}{}}  & \multirow{2}{*}{} &     &       &      &      \\[-2.3ex]
				\multicolumn{1}{l}{}                       &                           &      $\Gamma=0.31 $ GeV          &  $\Gamma=0.31 $ GeV          &     $\Gamma=0.12 $ GeV       &      $\Gamma=0.10 $ GeV        \\ \cmidrule{2-6}
				\multicolumn{1}{c}{\multirow{4}{*}{}}  & \multirow{2}{*}{} &     &       &      &      \\[-2.3ex]
				\multicolumn{1}{l}{}                       & \multirow{2}{*}{Model B}         &      NA   & NA        &       NA  &    $m=6.78 $ GeV          \\ 
				
				\multicolumn{1}{c}{\multirow{4}{*}{}}  & \multirow{2}{*}{} &     &       &      &      \\[-2.3ex]
				\multicolumn{1}{l}{}                       &               &   NA      &   NA      &   NA      &    $\Gamma=0.39 $ GeV          \\ \cmidrule{1-6}
				\multicolumn{2}{l}{\multirow{2}{*}{}}                                      &         &         &         &           \\[-2.3ex]
				\multicolumn{2}{l}{\multirow{2}{*}{CMS \cite{CMS}}}                                      &   NA      &   $m=6552$ MeV      & $m=6927$ MeV        &  $m=7287 $ MeV       \\ 
				\multicolumn{2}{l}{\multirow{2}{*}{}}                                      &     &         &         &          \\[-2.3ex]
				\multicolumn{2}{l}{}                                                          &     NA    &     $\Gamma=124$ MeV      &     $\Gamma=122$ MeV      &   $\Gamma=95$ MeV          \\ \bottomrule			
			\end{tabular}
		\end{center}
	\end{table*}
	In this work we will suggest possible quantum numbers for the $X(6900)$ together with its quark structure that explains seemingly lacking prominent ground state and create tables of possible masses of other all-charm and all-bottom tetraquarks applying a compact tetraquark framework.  The discussion will not involve in-medium effects. Numerical calculations have been performed using Wolfram Language and Mathematica 13 and the base code by F. Sch\"oberl  and W. Lucha  \cite{codeSource} which had been modified and adjusted to its application in the present work. The code can be found in the supplemental file 1. In this work we will use new notation for exotic hadrons suggested in reference \cite{ExoticNaming}.
	
	\section{Charmonia and bottomonia}
	
	The charmonium spectrum can be obtained through solving the time-independent Schr\"odinger equation with Fermi-Breit Hamiltonian \cite{BoundStates} with reduced mass $\mu_{12}=\frac{m_1 m_2}{m_1+m_2}$,
	\begin{subequations}
		\begin{equation}\label{Pair-Hamiltonian}
			\begin{split}
				\big[m_1+m_2+\frac{1}{2\mu_{12}}\big(-\frac{d^2}{dr^2} +\frac{l(l+1)}{r^2}\big)+ \\+V^S_{12}+V^{SS}_{12}+V^{LS}_{12}+V^T_{12}\big]\Psi=E^{n,l}_{12}\Psi.
			\end{split}
		\end{equation}
		The Hamiltonian includes kinetic energy and the contribution of strong interaction between the quark-antiquark pair. The pair interaction includes the one gluon exchange (OGE). For the charmonium and the bottomonium spectrum Cornell potential applies,
		\begin{equation}\label{Strong}
			V^G_{ij}(r_{ij})=\kappa_s\frac{\alpha_s}{r_{12}}+\sigma r_{12},
		\end{equation} along with three spin dependent terms; spin-spin term  $(V_{ij}^{SS})$, spin-orbit term $(V_{ij}^{SL})$ and tensor term $(V_{ij}^{T})$ ; 
		\begin{equation}\label{Spin-Spin}
			V_{ij}^{SS}(r_{ij})=-\frac{8\kappa_s\alpha_s\pi}{3m^2}(\frac{\sigma_{ss}}{\sqrt{\pi}})^3 e^{-\sigma_{ss}^2r_{ij}^2} S_iS_j,
		\end{equation}
		\begin{equation}\label{Spin-Orbit}
			V_{ij}^{LS}(r_{ij})=\big[-\frac{3\kappa_s \alpha_s}{2m^2}\frac{1}{r_{ij}^3}-\frac{b}{2m^2}\frac{1}{r_{ij}}\big] LS,
		\end{equation}
		\begin{equation}\label{Tensor}
			V_{ij}^{T}(r_{ij})=-\frac{12\kappa_s\alpha_s}{4m^2}\frac{1}{r_{ij}^3}\big(\frac{(S_ir_{ij})(S_jr_{ij})}{r_{ij}^2}-\frac{S_iS_j}{3}\big).
		\end{equation}
	\end{subequations}
	In the equation (\ref{Strong}) there are three parameters to be set.  The first one is the Casimir coefficient $\kappa_s$ that can be obtained by calculating scattering amplitude for the quark-antiquark pair that form a colour-singlet:
	\begin{equation}\label{Cassimircoof}
		\begin{split}
			\kappa_s^1=-f_1= \frac{1}{4}\Sigma_{\alpha}(c_3^\dagger \lambda^\alpha c_1)(c_2^\dagger \lambda^\alpha c_4)=\\
			=\frac{1}{4}\big(\frac{1}{\sqrt{3}}\big)^2 Tr(\lambda^\alpha \lambda^\alpha)=\frac{-4}{3}.
		\end{split}
	\end{equation}
	
	The other two parameters, strong coupling constant $\alpha_s$ and string tension $\sigma$ are obtained from a fit to the charmonium and bottomonium spectra and radiative decays \cite{Confinement}. Spin dependent terms include one additional parameter $\sigma_{ss}$ which is also impossible to determine empirically.
	Mesons are formed from two quarks with spin $\frac{1}{2}$ that can either form spin singlets ($S=0$) or spin triplets ($S=3$):
	\begin{equation}
		2\otimes2=1\oplus3.
	\end{equation}
	The contribution of the spin-spin internation in the $V^{SS}$ is evaluated using:
	\begin{equation}
		\langle S_i  S_j \rangle =\frac{1}{2}\langle S^2-S_1^2-S_2^2\rangle=\begin{cases}
			\frac{-3}{4},\quad  S=|0\rangle \\
			\frac{1}{4},\quad S=|1\rangle,
		\end{cases}
	\end{equation}
	where $S_1, S_2$ are the spins of each quark and $S$ is the total spin. 
	In $V^{SL}$ for $L\neq0$:
	\begin{equation}
		\langle LS\rangle_{S=|1\rangle}=
		\begin{cases}
			-(l-1), \quad J=L-1\\
			-1, \quad J=L\\
			l, \quad J=L+1.
		\end{cases}
	\end{equation}  
	The tensor dependent part of the potential (\ref{Tensor}) ought to be transformed with the usage of the identity (\ref{True-for-fermions}) \cite{TrueForFermions} to remove $r_{ij}$ dependency and include $J$ and $L$ instead (\ref{Tensor-decomp}).
	\begin{equation}\label{True-for-fermions}
		\begin{split}
			(a\cdot b)r^2-3(a \cdot r)(b\cdot r)=\frac{r^2}{(2l+3)(2l+1)}\times\\
			\times(k^2(a\cdot b)-3(a\cdot k)(b\cdot k)-3(b \cdot k)(a\cdot k))
		\end{split}
	\end{equation}
	\begin{equation}\label{Tensor-decomp}
		\begin{split}
			\langle T_{ij}\rangle=12\big\langle\frac{(S_ir_{ij})(S_jr_{ij})}{r_{ij}^2}-\frac{S_iS_j}{3}\big\rangle= \\
			=\frac{4}{(2l+3)(2l-1)}\big\langle S^2 L^2+\frac{3}{2}(LS)+3(LS)^2\big\rangle
		\end{split}.
	\end{equation}
	Therefore $V^T\neq0$ for $S\neq0$ and $L\neq0$:
	\begin{equation}
		\langle T_{ij}\rangle=
		\begin{cases}
			2, \quad J=L\\
			\frac{-2(l+1)}{2l-1}, \quad J=L-1\\
			\frac{-2l}{2l+3}, \quad J=L+1.
		\end{cases}
	\end{equation}
	We made appropriate choice of the parameters for charmonia (\ref{ParametersC}) and bottomonia (\ref{ParametersB}):
	\begin{subequations}
		\begin{equation}\label{ParametersC}
			\begin{split}
				\alpha_0=0.198, \quad \sigma=0.177\ \text{GeV}^2, \\
				\sigma_{ss}=1.08 , \quad
				m_c =1.263\ \text{GeV},
			\end{split}
		\end{equation}
		\begin{equation}\label{ParametersB}
			\begin{split}
				\alpha_0=0.164, \quad \sigma=0.177\ \text{GeV}^2, \\
				\sigma_{ss}=1.08,    \quad m_b =4.581\ \text{GeV}. 
			\end{split}
		\end{equation}
	\end{subequations}
	Masses and widths of the resonances were sourced from PDG \cite{ParticleDataGroup}. 
	
	\begin{table*}\captionof{table}{The $c\bar{c}$ mesons and their respective masses ($M_{exp}$) compared to the masses calculated in this work ($M_{cal}$) using parameters (\ref{ParametersC}).  The last column incudes the comparison of $M_{exp}$ with $M_{calc}$ in percentages.}\label{Table CCMesons}
		\begin{center}
			\begin{tabular}{ccccccc}
				\toprule
				\rule[-1ex]{0pt}{2.5ex} & & & & & & \\[-2.3ex]
				\rule[-1ex]{0pt}{2.5ex} Name & $N^{2S+1}l_J$ & $J^{PC}$ & $M_{exp}(\text{MeV})$ & $\Gamma(\text{MeV}) $ & $M_{cal}(\text{MeV})$ & $\frac{M_{cal}-M_{exp}}{M_{exp}} (\%)$ \\[0.6ex]
				\midrule
				\rule[-1ex]{0pt}{2.5ex} & & & & & & \\[-2.5ex]
				\rule[-1ex]{0pt}{2.5ex} $\eta_c(1S)$ & $1^1 S_0$ & $1^{-+}$ & $2983.9\pm0.5$ &$32.0 \pm0.7$ & 2983.8&  $< 0.1$ \\
				\rule[-1ex]{0pt}{2.5ex} & & & & & & \\[-2.5ex]
				\rule[-1ex]{0pt}{2.5ex} $J/\psi$ &$1^3 S_1$& $1^{--}$ & $3096.900\pm 0.006$ &  $92.9 \pm2.8$  &3094.7 &$< 0.1$ \\
				
				\midrule
				\rule[-1ex]{0pt}{2.5ex} & & & & & & \\[-2.5ex]
				\rule[-1ex]{0pt}{2.5ex}$h_c(1P)$  & $1^1P_1$  & $1^{+-}$ &$3525.38 \pm 0.11$  & $0.7 \pm 0.4$ & 3576.1 & 1.4\\
				\rule[-1ex]{0pt}{2.5ex} & & & & & & \\[-2.5ex]
				\rule[-1ex]{0pt}{2.5ex} $\chi_{c0}(1P)$ & $1^3 P_0$  & $0^{++}$ & $3414.71 \pm 0.30$ &$10.8 \pm 0.6$  & 3286.5 & 3.8\\
				\rule[-1ex]{0pt}{2.5ex} & & & & & & \\[-2.5ex]
				\rule[-1ex]{0pt}{2.5ex}$\chi_{c1}(1P)$  &$1^3 P_1$  & $1^{++}$ & $3510.67 \pm0.05$ &  $0.84 \pm 0.04$ & 3580.9&1.9\\
				\rule[-1ex]{0pt}{2.5ex} & & & & & & \\[-2.5ex]
				\rule[-1ex]{0pt}{2.5ex}$\chi_{c2}(1P)$  &$1^3 P_2$   & $2^{++}$ & $3556.17 \pm 0.07$  & $1.97 \pm 0.09$ & 3365.6& 5.4 \\
				
				\midrule
				\rule[-1ex]{0pt}{2.5ex} & & & & & & \\[-2.5ex]
				\rule[-1ex]{0pt}{2.5ex} $\eta_c(2S)$ & $2^1 S_0$ & $1^{-+}$ & $3637.5 \pm 1.1$ &$11.3 \pm 3.2$ & 3576.4 &  1.6\\
				\rule[-1ex]{0pt}{2.5ex} & & & & & & \\[-2.5ex]
				\rule[-1ex]{0pt}{2.5ex} $\psi(2S)$ & $2^3 S_1$ & $1^{--}$ & $3686.10 \pm 0.06$ &$294 \pm 8$  & 3641.9 & 1.2\\
				\rule[-1ex]{0pt}{2.5ex} & & & & & & \\[-2.5ex]
				\rule[-1ex]{0pt}{2.5ex} $\psi(3770)$ & $1^3 D_1$ & $1^{--}$ &$3773.7 \pm 0.4$  & $27.2 \pm 1.0$ & 3758.5 & 0.4\\
				\rule[-1ex]{0pt}{2.5ex} & & & & & & \\[-2.5ex]
				\rule[-1ex]{0pt}{2.5ex} $\psi_2(3823)*$ & $1^3 D_2$ & $2^{--}$ &$3823.7\pm 0.5$  & $< 5.2$ &3816.4 & 0.4\\
				\rule[-1ex]{0pt}{2.5ex} & & & & & & \\[-2.5ex]
				\rule[-1ex]{0pt}{2.5ex} $\psi_3(3842)*$& $1^3 D_3 $  &$3^{--}$  & $3842.71 \pm 0.20$ &  $2.8 \pm 0.6$ & 3602.8& 6.2\\
				
				\midrule
				\rule[-1ex]{0pt}{2.5ex} & & & & & & \\[-2.5ex]
				\rule[-1ex]{0pt}{2.5ex} $Z_c(3900)$ & $2^1P_1$  & $1^{+-}$  & $3887.1  \pm 2.6$ & $28.4 \pm 2.6$ & 4108.3& 5.7\\
				\rule[-1ex]{0pt}{2.5ex} & & & & & & \\[-2.5ex]
				\rule[-1ex]{0pt}{2.5ex} $\chi_{c0}(3915)$ & $2^3 P_0$  & $0^{++}$  & $3921.7 \pm 1.9$ & $18.8  \pm 3.5 $ & 3561.7 & 9.2\\
				\rule[-1ex]{0pt}{2.5ex} & & & & & & \\[-2.5ex]
				\rule[-1ex]{0pt}{2.5ex} $\chi_{c1}(3872)$ & $2^3 P_1$ &$1^{++}$  & $3871.65  \pm 0.06$ & $1.19\pm0.21$  & 4120.9& 6.4\\
				\bottomrule
			\end{tabular}
		\end{center}	
	\end{table*}
	\begin{table*}\captionof{table}{The $b\bar{b}$ mesons and their respective masses ($M_{exp}$) compared to the masses calculated in this paper ($M_{cal}$) using parameters (\ref{ParametersB}).  The last column incudes the comparison of $M_{exp}$ with $M_{cal}$ in percentages.}\label{Table BBMesons}
		\begin{center}
			\begin{tabular}{ccccccc}
				\toprule
				\rule[-1ex]{0pt}{2.5ex} & & & & & & \\[-2.3ex]
				\rule[-1ex]{0pt}{2.5ex} Name & $N^{2S+1}l_J$ & $J^{PC}$ & $M_{exp}(\text{MeV})$ & $\Gamma(\text{MeV}) $ & $M_{cal}(\text{MeV})$ & $\frac{M_{cal}-M_{exp}}{M_{exp}} (\%)$ \\[0.6ex]
				
				\midrule
				\rule[-1ex]{0pt}{2.5ex} & & & & & & \\[-2.5ex]
				\rule[-1ex]{0pt}{2.5ex} $\eta_b(1S)$ & $1^1 S_0$ & $1^{-+}$ & $9398.7\pm2.0$ &$10\pm5$ &9417.1  & 0.2   \\
				\rule[-1ex]{0pt}{2.5ex} & & & & & & \\[-2.5ex]
				\rule[-1ex]{0pt}{2.5ex} $\Upsilon(1S)$ &$1^3 S_1$& $1^{--}$ & $9460.30\pm0.26$ &  $\approx0.054\pm0.001$  & 9397.4 & 0.7 \\
				
				\midrule
				\rule[-1ex]{0pt}{2.5ex} & & & & & & \\[-2.5ex]
				\rule[-1ex]{0pt}{2.5ex} $\chi_{b0}(1P)$ & $1^3 P_0$  & $0^{++}$ & $9859.44 \pm 0.26  $ &NA   & 9417.1 & 4.4\\
				\rule[-1ex]{0pt}{2.5ex} & & & & & & \\[-2.5ex]
				\rule[-1ex]{0pt}{2.5ex}$\chi_{b1}(1P)$  &$1^3 P_1$  & $1^{++}$ & $9892.78 \pm0.05$ &  NA& 9708.2 & 1.9\\
				\rule[-1ex]{0pt}{2.5ex} & & & & & & \\[-2.5ex]
				\rule[-1ex]{0pt}{2.5ex}$\chi_{b2}(1P)$  &$1^3 P_2$   & $2^{++}$ & $9912.21 \pm 0.26$  & NA & 9674.6& 2.3 \\
				
				\midrule
				\rule[-1ex]{0pt}{2.5ex} & & & & & & \\[-2.5ex]
				\rule[-1ex]{0pt}{2.5ex} $\Upsilon(2S)$ & $2^3 S_1$ & $1^{--}$ & $10023.26\pm 0.31 $ &$\approx 0.031 \pm 0.003$  & 9814.6 & 2.1\\
				\rule[-1ex]{0pt}{2.5ex} & & & & & & \\[-2.5ex]
				\rule[-1ex]{0pt}{2.5ex} $\Upsilon(3S)$ & $2^3 S_1$ & $1^{--}$ & $10355.2 \pm 0.5 6$ &$\approx 0.020 \pm 0.002$  & 10102.4 & 2.4\\
				\rule[-1ex]{0pt}{2.5ex} & & & & & & \\[-2.5ex]
				\rule[-1ex]{0pt}{2.5ex} $\Upsilon(4S)$ & $2^3 S_1$ & $1^{--}$ & $10579.4 \pm 1.2 $ &$20.5 \pm 2.5$  & 10358.4 & 1.3\\
				\midrule
				\rule[-1ex]{0pt}{2.5ex} & & & & & & \\[-2.5ex]
				\rule[-1ex]{0pt}{2.5ex} $\Upsilon_2(1D)$ & $1^3 D_2$ & $2^{--}$ &$10163.7 \pm 1.4$  & NA & 10062.4 & 1.0\\
				\midrule
				\rule[-1ex]{0pt}{2.5ex} & & & & & & \\[-2.5ex]
				\rule[-1ex]{0pt}{2.5ex} $h_b(1P)$ & $1^1P_1$  & $1^{+-}$  & $9899.3 \pm 0.8  $ & NA & 9713.0& 1.9\\
				\bottomrule
			\end{tabular}		
		\end{center}
	\end{table*}
	
	\section{Tetraquarks}
	The tools that were first used to properly describe non-exotic hadrons \cite{GellMann}\cite{GellMann2} may be applied to the tetraquarks and other exotic hadrons \cite{Jaffe1}\cite{Jaffe2}. The discussion of potential tetraquark states predates the discovery of the first exotic hadron candidate $\chi_{c1}(3872)$, previously known as $X(3872)$, in 2003 \cite{BelleX3872}. In terms of the fundamental representation, a quark and an antiquark could form a colour-singlet meson that can be observed in the resonances or a colour octet (\ref{1+8}). Non-singlet colour states have not been observed \cite{ParticleDataGroup} due to the colour confinement, however we can treat newly build colour octets as our new "building blocks" that may be combined into a colour singlet and multiple non-singlets (\ref{8x8}).  
	\begin{subequations}
		\begin{equation} \label{1+8}
			3\otimes\bar{3}=1\oplus8
		\end{equation}
		\begin{equation}\label{8x8}
			8\otimes8=1\oplus8\oplus8\oplus10\oplus\bar{10}\oplus27
		\end{equation}
		The concept of a diquark has been frequently used to describe exotic hadrons since the conception of the quark model \cite{Diquarks-rev}. Compact tetraquarks are usually described as bound states of a diquark and an antidiquark forming an exotic meson \cite{MultiquarkHadrons}. In this work we will call this configuration "diquark-antidiquark configuration" or "DA". In addition to the diquark picture we will also consider combination of two colour octets, referred to as "meson-meson configuration" or "MM". To obtain masses of the tetraquarks we use the parameters calculated in the previous section to calculate masses of colour non-singlet states.  
		\begin{equation}
			3\otimes3=\bar{3}\oplus6, \quad \bar{3}\otimes\bar{3}=3\oplus\bar{6}
		\end{equation}
		\begin{equation}
			6\otimes\bar{6}=1\oplus8\oplus27
		\end{equation}
		We can see that there are several possible methods of "composing" tetraquaks. Using the same method as in Eq. (\ref{Cassimircoof}) to obtain Casimir Coefficient for colour singlet, we archive the results displayed in the Table \ref{Cass} and Table \ref{Cass2}. The parameters used in the  Cornell Potential Eq. (\ref{Strong}) for interactions between the pairs are identical for those used to obtain charmonium and bottomonium spectra, modified by the Casimir scaling (\ref{Scaling}) \cite{Bali:2000un}\cite{ScalingDeldar:1999vi}. A transition from the DA to the MM configuration could be performed unless the distance between the diquark and the antidiquark is sufficiently larger than the distance between their constituents \cite{Gromes-Transfer}, therefore the parameters $\alpha_s$ and $\sigma$ need to be scaled accordingly.
	\end{subequations}
	
	\begin{table}[h]\captionof{table}{The Casimir coefficients for the various colour structures of two partons.}\label{Cass}
		\begin{tabular}{llllll}
			\toprule
			& \multicolumn{2}{l}{$q\bar{q}$} & & \multicolumn{2}{l}{$qq$} \\ \cmidrule(l){2-3} \cmidrule(l){5-6}
			\multicolumn{1}{l}{} &           &   & &        &       \\[-2.3ex]
			\multicolumn{1}{l}{Colour state} &   $  1$      & $  8 $  &     &    $ \bar{3}$      &   $  6 $     \\ 
			\multicolumn{1}{l}{} &           &    &    &    &       \\[-2.3ex] 
			\multicolumn{1}{l}{$\kappa_s$} &     $-4/3$      &   $ 1/6$    &   &   $-2/3$        &  $1/3$         \\ \bottomrule
		\end{tabular}
	\end{table}
	\begin{table}[h]\captionof{table}{Casimir coefficients for
			the two cluster interaction.}\label{Cass2}
		\begin{tabular}{llllll}
			\toprule
			& \multicolumn{2}{l}{} & \multicolumn{2}{l}{} \\[-2.3ex]
			& \multicolumn{2}{l}{$MM$} & &\multicolumn{2}{l}{$D\bar{D} $} \\\cmidrule(l){2-3} \cmidrule(l){5-6}
			\multicolumn{1}{l}{} &           &&    &        &       \\[-2.3ex]
			\multicolumn{1}{l}{Colour state} &     $  MM_{11}$     &   $MM_{88}$      &   & $ DA_{3\bar{3}}$      &     $DA_{6\bar{6}}$   \\[0.2ex]
			\multicolumn{1}{l}{} &           &    &        &      & \\[-2.3ex]
			\multicolumn{1}{l}{$\kappa_s$} &     $-4/3$      &   $ -3$       &&   $-4/3$        &  $-10/3$    \\\bottomrule
		\end{tabular}
	\end{table}
	\begin{equation}\label{Scaling}
		\frac{V_1(r)}{V_r(2)}=\frac{\kappa_{s1}}{\kappa_{s2}}
	\end{equation} 
	The compact tetraquark picture is a description in which interaction between a diquark and an antidiquark is treated similarly to interaction between a quark and an antiquark in a non-exotic meson \cite{AAA}. The method of using diquarks in the description of hadronic bound states is not unique to tetraquarks as it can be used for baryons \cite{Blaschke}\cite{BARYON-Basdevant} and pentaquarks \cite{Jaffe-diquarks}\cite{Pentaquarks}. Using the numerical method described in previous section we will calculate masses of the diquarks and the non-singlet quark-antiquark states and use the obtained result to calculate masses of the tetraquarks composed of those pairs, or in other words, we treat the problem like a set of three different two-body problems. In this work we will use Columb-like potential to describe boson exchange between clusters. 
	
	\begin{table*}[h]\captionof{table}{Masses of colour sextet diquarks for possible combinations of their quantum numbers.}\label{TableSextet}
		\begin{center}
			\begin{tabular}{lllllllll}
				\toprule
				\multirow{3}{*}{$\kappa_s$=$\frac{1}{3}$}&  & \multicolumn{2}{c}{S=0} &  & \multicolumn{4}{c}{S=1}        \\\cmidrule{3-4}\cmidrule{6-9}
				&  & L=0        & L=1        &  & L=0  & \multicolumn{3}{c}{L=1} \\\cmidrule(l){3-3}\cmidrule(l){4-4}\cmidrule(l){6-6}\cmidrule(l){7-9}
				&  & J=0        & J=1        &  & J=1  & J=0     & J=1   & J=2   \\\midrule
				N=1 &  & 2819.4          &  2929.4         &  & 2815.1     &  2997.3       & 2929.2      &  2912.0     \\
				N=2 & & 3016.0 &  3100.7           &  &    3011.9  &     3188.6    &  3100.3     & 3084.1  \\ \bottomrule  
			\end{tabular}
		\end{center}
	\end{table*}
	\begin{table*}[h]\captionof{table}{Masses of colour triplet diquarks for possible combinations of their quantum numbers.} \label{TableTriplet}
		\begin{center}
			\begin{tabular}{lllllllll}
				\toprule
				\multirow{3}{*}{$\kappa_s$=$-\frac{2}{3}$}&  & \multicolumn{2}{c}{S=0} &  & \multicolumn{4}{c}{S=1}        \\\cmidrule{3-4}\cmidrule{6-9}
				&  & L=0        & L=1        &  & L=0  & \multicolumn{3}{c}{L=1} \\\cmidrule(l){3-3}\cmidrule(l){4-4}\cmidrule(l){6-6}\cmidrule(l){7-9}
				&  & J=0        & J=1        &  & J=1  & J=0     & J=1   & J=2   \\\midrule
				N=1 &  & 2882.3         &  3172.6       &  & 2908.8     &  3072.3       & 3173.8      &  3083.6     \\
				N=2 & & 3226.2 &  3503.8           &  &    3245.6 &     3263.1    &  3503.2      & 3363.5    \\ \bottomrule
			\end{tabular}
		\end{center}
	\end{table*}
	\begin{table*}[t]\captionof{table}{Masses of colour octet quark-antiquark states for possible combinations of their quantum numbers.}\label{TableOctet}
		\begin{center}
			\begin{tabular}{lllllllll}
				\toprule
				\multirow{3}{*}{$\kappa_s$=$\frac{1}{6}$}&  & \multicolumn{2}{c}{S=0} &  & \multicolumn{4}{c}{S=1}        \\\cmidrule{3-4}\cmidrule{6-9}
				&  & L=0        & L=1        &  & L=0  & \multicolumn{3}{c}{L=1} \\\cmidrule(l){3-3}\cmidrule(l){4-4}\cmidrule(l){6-6}\cmidrule(l){7-9}
				&  & J=0        & J=1        &  & J=1  & J=0     & J=1   & J=2   \\\midrule
				N=1 &  & 2705.2          &  2777.3        &  & 2704.0     &  2808.8       & 2777.2      &  2768.6     \\
				N=2 & & 2830.4 &  2886.9           &  &    2829.3  &     2922.4    &  2886.8     & 2878.3   \\ \bottomrule 
			\end{tabular}
		\end{center}
	\end{table*}
	
\begin{figure}[h]
		\centering
		\includegraphics[width=0.85\linewidth]{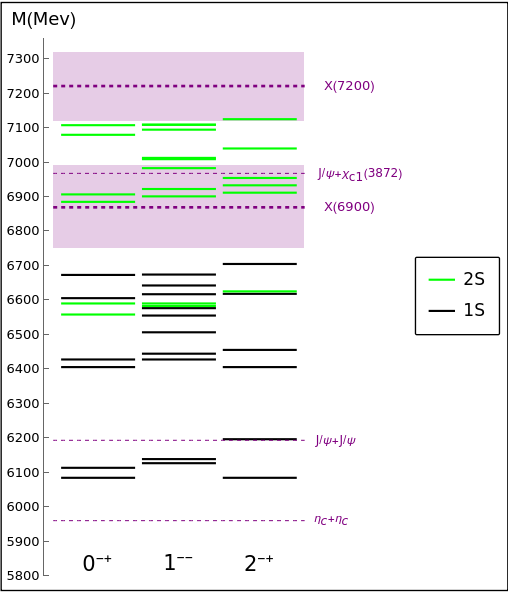}
		\caption{Energy level diagram for $DA_{3\bar{3}}$ all-charm tetraquark structure. Numerical data can be found in Table \ref{ResultsTable1} and Table \ref{ResultsTable2}. Masses and widths of the resonances $X(6900)$ and $X(7200)$ were obtained by ATLAS with model A \cite{ATLAS2022}.}
		\label{fig:plotactc1}
	\end{figure}

\begin{figure}[h]
		\centering
		\includegraphics[width=0.85\linewidth]{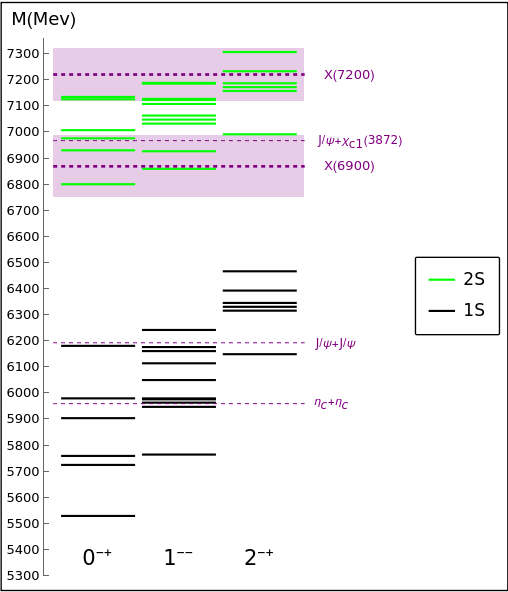}
		\caption{Energy level diagram for $DA_{6\bar{6}}$ all-charm tetraquark structure. Numerical data can be found in Table \ref{ResultsTable1} and Table \ref{ResultsTable2}. Masses  and widths of the resonances $X(6900)$ and $X(7200)$ were obtained by ATLAS with model A \cite{ATLAS2022}.}
		\label{fig:plotactc2}
	\end{figure}

\begin{figure}[h]
		\centering
		\includegraphics[width=0.85\linewidth]{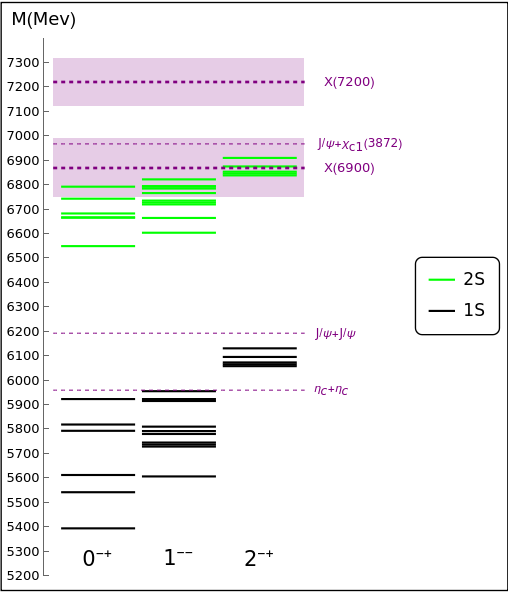}
		\caption{Energy level diagram for $MM_{88}$ all-charm tetraquark structure. Numerical data can be found in Table \ref{ResultsTable1} and Table \ref{ResultsTable2}. Masses  and widths of the resonances $X(6900)$ and $X(7200)$ were obtained by ATLAS with model A \cite{ATLAS2022}.}
		\label{fig:plotactc3}
	\end{figure}

	The Hamiltonian used to calculate masses of the tetraquarks is presented in  Eq. (\ref{Tetra-Hamiltonian}) with $H_T$ representing contribution of kinetic energy and $(V_{ij})$ representing all the potential contributions including the spin contributions.
	\begin{subequations}
		\begin{equation}\label{Tetra-Hamiltonian}
			\begin{split}
				H=H_T+V_{(12)(34)}+\\
				+(m_1+m_2+V_{12})+(m_3+m_4+V_{34})
			\end{split}
		\end{equation}
		\begin{equation}\label{Kinetic-Ham}
			H_T=\frac{\hat{p}_{12}^2}{2\mu_{12}}+\frac{\hat{p}_{34}^2}{2\mu_{34}}+\frac{\hat{p}_{1234}^2}{2\mu_{1234}}
		\end{equation}
		\begin{equation}\label{Red-Mas1}
			\begin{split}
				\mu_{12}=\frac{m_1 m_2}{m_1+m_2}, \quad  	\mu_{34}=\frac{m_3 m_4}{m_3+m_4}\\
				\mu_{1234}=\frac{\mu_{12} \mu_{34}}{\mu_{12}+\mu_{34}}
			\end{split}
		\end{equation}
		\begin{equation}
			\begin{split}
				\vec{r}_{12}=\vec{r}_1-\vec{r}_2, \quad \vec{r}_{34}=\vec{r}_3-\vec{r}_4 \\
				\vec{r}_{1234}=\vec{r}_{34}-\vec{r}_{12}
			\end{split}
		\end{equation}
		\begin{equation}
			\begin{split}
				\vec{R}_{12}=\frac{m_1 r_1}{m_1+m_2}+\frac{m_2 r_2}{m_1+m_2}, \\ \vec{R}_{34}=\frac{m_3 r_3}{m_3+m_4}+\frac{m_4 r_4}{m_3+m_4}
			\end{split}
		\end{equation}
		\begin{equation}
			\vec{R}_{1234}=\frac{\mu_{12} r_{12}}{\mu_{12}+\mu_{34}}+\frac{\mu_{34} r_{34}}{\mu_{12}+\mu_{34}}
		\end{equation}	
	\end{subequations} 
	One of the significant differences between classic meson and compact tetraquark is the fact that pairs are bosons, therefore we cannot use identity found in Eq. (\ref{True-for-fermions}). Complete representation can be written as:
	\begin{equation}
		(2\otimes2)\otimes(2\otimes2)=(1\oplus3)\otimes(1\oplus3)=1\oplus3\oplus3\oplus1\oplus3\oplus5,
	\end{equation}
	therefore for  the pair-pair interaction in (\ref{Spin-Spin}):
	\begin{equation}
		\begin{split}
			\langle S_{12}  S_{34} \rangle =
			\frac{1}{2}\langle S^2-S_1^2-S_2^2\rangle=\\
			=\begin{cases}
				0, \quad  S=|0\rangle, S_{12}=S_{34}=|0\rangle \\
				0, \quad  S=|1\rangle, S_{12}\neq S_{34} \\
				-2, \quad S=|0\rangle, S_{12}=S_{34}=|1\rangle \\
				-1, \quad S=|1\rangle, S_{12}=S_{34}=|1\rangle \\
				1, \quad  S=|2\rangle, S_{12}=S_{34}=|1\rangle \\
			\end{cases}.
		\end{split}
	\end{equation}
	Because all of the resonances were identified in $di-J/\psi$ channel (except for $X(7200)$ which was additionally identified in $J/\psi+\Psi(2S)$ \cite{ATLAS2022}) we will assume that total angular momentum is $L=0$, so $V^{SL}$ (\ref{Spin-Orbit}) and $V^T$ (\ref{Tensor}) can be neglected.  Additional adjustment should be made to the $\alpha_s$ to include its scale dependence. Relation (\ref{running}) \cite{Running}\cite{Running-Buchmuller:1980su} with parameters (\ref{runningparamters}) and $\mu$ as the reduced mass of the system. 
	\begin{subequations}
		\begin{equation}\label{running}
			\alpha_s(\mu)=\frac{\alpha_0}{ln(\frac{\mu^2+\mu_0^2}{\Lambda_0^2})}
		\end{equation} 
		\begin{equation}\label{runningparamters}
			\mu_0 \approx0\ \text{MeV},\quad \Lambda_0=0.112\ \text{MeV}, \quad \alpha_0=3.2524
		\end{equation}
	\end{subequations}
	For the purposes of this work we will use following naming scheme for potential tetraquark candidates that includes two-letter code, "$MM$"  or  "$DA$", with subscript denting possible colour configuration and three sets of quantum terms; one for total system, two for the pairs. An example of the usage this naming scheme can be found in Eq. (\ref{Naming}.) The naming scheme can potentially be used to label wave functions for those states. 
	\begin{equation}\label{Naming}
		DA_{3\bar{3}}(2^3S_1\leftarrow1^1L_1+1^1 L_0).
	\end{equation} 
	Results with masses of the diquarks have been calculated with use of the code and placed in the Tables \ref{TableSextet}-\ref{TableOctet}. Masses of the tetraquarks were calculated by choosing quantum numbers of the components, quantum numbers of their product and possible colour combination. The code allows to consider other types of interaction than the Cornell potential or to consider massive force carriers, however, we want to stay on safe ground concerning the heavy quarkonia spectroscopy. The exact results for the ground state ($N=0$) and the first orbitral excitation ($N=1$) can be for in the tables and energy level diagrams can be found in Figs \ref{fig:plotactc1}-\ref{fig:plotactc3}.
	
	\section{Discussion of the results}
	
	We can clearly see that regardless of the composition of the tetraquark there are multiple candidates for for $X(6900)$. All of the possible candidates are $N=2$ states. Only sextet-antisextet states have suitable $X(7200)$ candidates and all the suitable $X(6900)$ candidates in that configuration have components with $L=0$. Due to the use of a simple model one definite configuration cannot be suggested with certainty, however the closest lying candidate is $DA_{6\bar{6}}(2^3S_1\leftarrow1^3 S_1+1^3 S_1)$. 
	In the previous section the subject of the wave function was omitted, as it held no significant impact on the calculation. However, we have to consider influence of the Pauli exclusion principle on the possibility of the existence of several resonances. Using the example shown in the Eq. (\ref{Naming}) we will show the use of the notation we introduced to write down the wave function of the tetraquark:
	\begin{equation}
		\begin{split}
			|DA_{3\bar{3}}(2^3S_1\leftarrow1^1L_1+1^1 L_0)\rangle=|D_{\bar{3}}(1^1 L_1)\rangle\langle A_{3}(1^1 L_1)|,\\
			|X_{c}(N^{2S+1} l_J)\rangle = |Y^m_l \times X^\sigma \times \ X^c \times X^f\rangle,
		\end{split}
	\end{equation}
	The four parts of the wave wave function are the spherical harmonics $Y^m_l$, the spin wave function $X^\sigma$, the colour wave function $X^c$ and the flavour wave function $X^f$.
	The flavour wave functions $X^f$ of a diquark $|D\rangle $ or antiquark $|A\rangle $ is symmetric, unlike the the wave function for a meson or a meson-like state $|M\rangle$;
	\begin{subequations}
		\begin{equation}
			\begin{split}
				|X^f_D\rangle=|cc\rangle, \quad |X^f_A\rangle=|\bar{c}\bar{c}\rangle, \quad |X^f_M\rangle= |c\bar{c}\rangle.
			\end{split}
		\end{equation}
		The spin wave function is identical for any pair and the total spin wave function $X^\sigma_{S, S_z}$ is dependent on the spin wave functions of the pairs:
		\begin{equation}
			\begin{split}
				|X^\sigma_{0,0} \rangle= \frac{1}{\sqrt{2}}(|\uparrow\downarrow\rangle - |\downarrow\uparrow\rangle), \\	|X^\sigma_{1,0}\rangle=\frac{1}{\sqrt{2}}( |\uparrow\downarrow\rangle + |\downarrow\uparrow\rangle),\\
				|X^\sigma_{1,1} \rangle= |\uparrow\uparrow\rangle, \qquad |X^\sigma_{1,-1}\rangle= |\downarrow\downarrow\rangle.
			\end{split}
		\end{equation}
		For the pairs only the $|X_{0,0}\rangle$ is antisymmetric. 
		The colour wave functions for the pairs will not be shown explicitly for the sake of brevity, however it needs to be explicitly stated that $|X^c_3\rangle$ is fully antisymmetric while $|X^c_6\rangle$ is fully symmetric.  
	\end{subequations}
	
	The wave function for the colour antisextet $1^3 S_1$ diquark would have a wave function that would violate the Pauli exclusion principle, therefore any state with that subsystem is marked as forbidden. 
	
	\begin{subequations}
		We have only two possible flavour wave functions for the tetraquark:
		\begin{equation}
			\begin{split}
				|X^f_{MM}\rangle = |X^f_{M}\rangle\langle X^f_{M}| = |c\bar{c} c\bar{c} \rangle, \\
				|X^f_{DA}\rangle = |X^f_{D}\rangle\langle X^f_{A}| = |cc\bar{c} \bar{c} \rangle.
			\end{split}
		\end{equation}
		Assuming that the total spin of the tetraquark is known as well as the spin of the both pairs we can obtain the total of six possible total spin wave functions:
		\begin{equation}
			\begin{split}
				|X_{0,0}^{\sigma_1}\rangle = |X^\sigma_{0,0}\rangle\langle X^\sigma_{0,0}|\\
				|X_{0,0}^{\sigma_2}\rangle = \frac{1}{\sqrt{3}}(|X^\sigma_{1,1}\rangle\langle X^\sigma_{1,-1}|+|X^\sigma_{1-,1}\rangle\langle X^\sigma_{1,1}|\\
				-|X^\sigma_{1,0}\rangle\langle X^\sigma_{1,0}|)\\
				|X_{1,1}^{\sigma_3}\rangle = |X^\sigma_{1,1}\rangle\langle X^\sigma_{0,0}|\\
				|X_{1,1}^{\sigma_4}\rangle = |X^\sigma_{0,0}\rangle\langle X^\sigma_{1,1}|\\
				|X_{1,1}^{\sigma_5}\rangle =\frac{1}{\sqrt{2}}( |X^\sigma_{1,1}\rangle\langle X^\sigma_{1,0}|+|X^\sigma_{1,0}\rangle\langle X^\sigma_{1,1}|)\\
				|X_{2,2}^{\sigma_6}\rangle = |X^\sigma_{1,1}\rangle\langle X^\sigma_{1,1}|.
			\end{split}
		\end{equation}
		There are only three total colour wave functions that we need to consider:
		\begin{equation}
			\begin{split}
				|X^c_{88}\rangle=|X^c_8\rangle\langle X^c_8|,\\
				|X^c_{3\bar{3}}\rangle=|X^c_{\bar{3}}\rangle\langle X^c_3|, \quad|X^c_{6\bar{6}}\rangle=|X^c_6\rangle\langle X^c_{\bar{6}}|.
			\end{split}
		\end{equation}
	\end{subequations}
	
	For the $L=0$ the spherical harmonic $Y^0_0$ is symmetric therefore $ X^\sigma \times \ X^c \times X^f$ must be antisymmetric, therefore we can mark $DA_{3\bar{3}}(1^1 S_0 \leftarrow1^1 S_0+1^1 S_0)$ as forbidden. 
	
	
	One of the problems that we are attempting to solve in this work is finding the reason of the high mass of the $X(6900)$ and the lack of more prominent lighter ground state. If we assume that the resonance has $DA_{6\bar{6}}$ structure we could easily explain lack of a prominent  resonances closer to $di - J/\psi$ and $di - \eta_c$ mass threshold could be explained by them violating Pauli exclusion principle, while states with $N=2$ could possibly be observed.   
	\begin{figure}[h]
		\centering
		\includegraphics[width=0.99\linewidth]{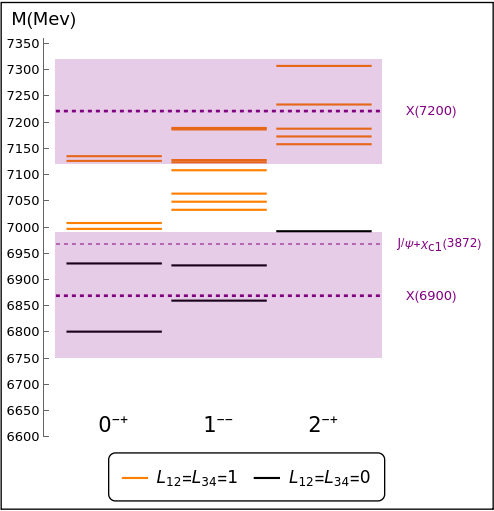}
		\caption{Energy level diagram for $DA_{6\bar{6}}$ and $N=2$ separated by orbital angular momentum of the substructures. Masses and widths of the resonances $X(6900)$ and $X(7200)$ were obtained by ATLAS with model A \cite{ATLAS2022}.}
		\label{fig:plotactc4}
	\end{figure}
	Further inspection of the data suggest that $X(6900)$ most likely is either $0^{-+}$ or $1^{--}$ state that have substructures with orbital angular momentum $L_{12}=L_{34}=0$. No suggestion for $X(7200)$ can be made, however if we assume that it has the same colour composition as $X(6900)$ we could assume that it state with $N=2$ and $L_{12}=L_{34}=1$. The possibility of $X(7200)$ being a $N=3$ state was not evaluated due to a lack of precision during the calculation. The other possible compositions of all-charm tetraquarks might be contributing to the broad structure. Since the background consists of multiple components and is possibly composed of several all-charm states with different quantum numbers, the confirmation of quantum numbers of $X(6900)$ and the other structures would require extraordinary precision and the ability to separate data from multiple hadrons with identical content. 
	
	Assuming that the $bb\bar{b}\bar{b}$ would behave in the same manner as $cc\bar{c}\bar{c}$, we can calculate potential mass of all-bottom tetraquark. Assuming that the composition of $DA_{6\bar{6}}(2^3S_1\leftarrow1^3 S_1+1^3 S_1)$ we archive:
	\begin{equation}
		m(bb\bar{b}\bar{b})\approx19.23\ \text{GeV}.
	\end{equation} 
	Due to the high mass, it might be unlikely to confirm the existence of the resonance in the next few years. 
	
	Investigation of the additional decay channels may be prolific, in particular investigation of channel involving another exotic hadron and/or D mesons. Despite the lack of significant data about some of the XYZ states, we have hypotheses regarding their internal dynamics, therefore gaining more information about possibility of the investigation of intermediate decays may possibly bring more information about tetraquark structure. Possible candidates for decay channels are $X(6900)\rightarrow\chi_{c1}(3872)+J/\psi$ or  $X(6900)\rightarrow \chi_{c1}(3872)+\eta_c$. State $\chi_{c1}(3872)$ is one of the lowest-lying exotic states and potential identification of it in the decays may help with obtaining the threshold from sharp all-charm resonances, since mass of $X(6900)$ is approximately equal to the sum of masses of the ground state charmonium and $\chi_{c1}(3872)$. 
	
	\section{Conclusions}
	
	Assuming  $L=0$,  the $X(6900)$ might have sextet-antisextet colour structure with $N=1$  and possible terms $0^{-+}$ or $1^{--}$, possibly a mixed state of both. The lack of a  more prominent resonance with lower energy could be explained by Pauli exclusion principle. Some of the low-lying $DA_{6\bar{6}}$ resonances do not appear as sharp peaks due to the impact of the non-zero angular momenta and due to the large number of the states with similar energies, instead they contribute to the broad structures around 6200 MeV and 6500 MeV. 
	
	The octet-octet and triplet-antitriplet structures contribute to the background and to the low-lying broad structures. One could consider the possibility that the $DA_{3\bar{3}}$ structure is less prominent than $DA_{6\bar{6}}$ due to a higher likelihood of the formation of a doubly-charmed baryon and and its antiparticle instead of tightly bound tetraquark. 
	More experimental data is needed to determine the quantum numbers of $X(6900)$ and investigating the $J/\psi+\Psi(2S)$ invariant mass spectrum to more precisely determine the mass of $X(7200)$ and the quantum numbers of the resonances. Additional decay channels involving the XYZ states, in particular  $\chi_{c1}(3872)$ should be considered. 
	
	\subsection{Acknowledgement}
		I would like to express my gratitude to my supervisor, prof. dr hab. David Blaschke for the support during my thesis research and writing of this paper.

	

	\begin{table*}[]\captionof{table}{Possible masses of the all-charm tetraquark for $N=1$. States marked with * are forbidden by the Pauli exclusion principle.}\label{ResultsTable1}
		\begin{center}
			\begin{tabular}{llllll}
				\toprule
				&                        &   & \multicolumn{3}{l}{}                            \\[-2.4ex] 
				&                        &   & \multicolumn{3}{l}{$M_{12,34}$ (MeV)}                            \\[0.4ex] \cmidrule{4-6}
				\multicolumn{1}{l}{} & \multicolumn{1}{l}{} &  & \multicolumn{1}{l}{} & \multicolumn{1}{l}{} &  \\[-2.4ex]
				\multicolumn{1}{l}{$(N^{2S+1} l_J)_{12}$} & \multicolumn{1}{l}{$(N^{2S+1} l_J)_{34}$} & $(N^{2S+1} l_J)_{12,34}$ & \multicolumn{1}{l}{$MM_{88}$} & \multicolumn{1}{l}{$DA_{6\bar{6}}$} & $DA_{3\bar{3}}$ \\[0.4ex] \midrule			\multicolumn{1}{l}{}  & \multicolumn{1}{l}{}  &   & \multicolumn{1}{l}{}   & \multicolumn{1}{l}{}   &    \\[-2.45ex]
				\multicolumn{1}{l}{$1^1 S_0$}  & \multicolumn{1}{l}{$1^1 S_0$}  &  $1^1 S_0$ & \multicolumn{1}{l}{5794.3}   & \multicolumn{1}{l}{5980.8}   &  6114.0*  \\ 
				\multicolumn{1}{l}{}  & \multicolumn{1}{l}{}  &   & \multicolumn{1}{l}{}   & \multicolumn{1}{l}{}   &    \\[-2.45ex]
				\multicolumn{1}{l}{$1^1 S_0$}  & \multicolumn{1}{l}{$1^3 S_1$}  & $1^3 S_1$  & \multicolumn{1}{l}{5793.2}   & \multicolumn{1}{l}{5977.1*}   &    6139.3\\ 
				\multicolumn{1}{l}{}  & \multicolumn{1}{l}{}  &   & \multicolumn{1}{l}{}   & \multicolumn{1}{l}{}   &    \\[-2.45ex]
				\multicolumn{1}{l}{$1^3 S_1$}  & \multicolumn{1}{l}{$1^3 S_1$}  & $1^1 S_0$  & \multicolumn{1}{l}{5395.0}   & \multicolumn{1}{l}{5530.2*}   &  6085.0 \\ 
				\multicolumn{1}{l}{}  & \multicolumn{1}{l}{}  &   & \multicolumn{1}{l}{}   & \multicolumn{1}{l}{}   &    \\[-2.45ex]
				\multicolumn{1}{l}{$1^3 S_1$}  & \multicolumn{1}{l}{$1^3 S_1$}  & $1^3 S_1$  & \multicolumn{1}{l}{5607.5}   & \multicolumn{1}{l}{5765.4*}   &  6127.4 \\ 
				\multicolumn{1}{l}{}  & \multicolumn{1}{l}{}  &   & \multicolumn{1}{l}{}   & \multicolumn{1}{l}{}   &    \\[-2.45ex]
				\multicolumn{1}{l}{$1^3 S_1$}  & \multicolumn{1}{l}{$1^3 S_1$}  & $1^5 S_2$  & \multicolumn{1}{l}{5946.2}   & \multicolumn{1}{l}{6150.1*}   &  6197.0  \\ \midrule

				\multicolumn{1}{l}{}  & \multicolumn{1}{l}{}  &   & \multicolumn{1}{l}{}   & \multicolumn{1}{l}{}   &    \\[-2.45ex]
				\multicolumn{1}{l}{$1^1 L_1$}  & \multicolumn{1}{l}{$1^1 L_1$}  &  $1^1 S_0$ & \multicolumn{1}{l}{5924.2}   & \multicolumn{1}{l}{6182.0}   &   6673.8 \\ 
				\multicolumn{1}{l}{}  & \multicolumn{1}{l}{}  &   & \multicolumn{1}{l}{}   & \multicolumn{1}{l}{}   &    \\[-2.45ex]
				\multicolumn{1}{l}{$1^1 L_1$}  & \multicolumn{1}{l}{$1^3 L_0$}  & $1^3 S_1$  & \multicolumn{1}{l}{5956.4}   & \multicolumn{1}{l}{6243.0}   &    6577.4\\ 
				\multicolumn{1}{l}{}  & \multicolumn{1}{l}{}  &   & \multicolumn{1}{l}{}   & \multicolumn{1}{l}{}   &    \\[-2.45ex]
				\multicolumn{1}{l}{$1^1 L_1$}  & \multicolumn{1}{l}{$1^3 L_1$}  & $1^3 S_1$  & \multicolumn{1}{l}{5924.2}   & \multicolumn{1}{l}{6177.4}   &  6674.9\\ 
				\multicolumn{1}{l}{}  & \multicolumn{1}{l}{}  &   & \multicolumn{1}{l}{}   & \multicolumn{1}{l}{}   &    \\[-2.45ex]
				\multicolumn{1}{l}{$1^1 L_1$}  & \multicolumn{1}{l}{$1^3 L_2$}  & $1^3 S_1$ & \multicolumn{1}{l}{5916.3}   & \multicolumn{1}{l}{6162.0}   & 6617.8 \\ \midrule

				\multicolumn{1}{l}{}  & \multicolumn{1}{l}{}  &   & \multicolumn{1}{l}{}   & \multicolumn{1}{l}{}   &    \\[-2.45ex]
				\multicolumn{1}{l}{$1^3 L_0$}  & \multicolumn{1}{l}{$1^3 L_0$}  &  $1^1 S_0$ & \multicolumn{1}{l}{5613.4}   & \multicolumn{1}{l}{5904.9}   &   6406.3 \\ 
				\multicolumn{1}{l}{}  & \multicolumn{1}{l}{}  &   & \multicolumn{1}{l}{}   & \multicolumn{1}{l}{}   &    \\[-2.45ex]
				\multicolumn{1}{l}{$1^3 L_0$}  & \multicolumn{1}{l}{$1^3 L_0$}  &  $1^3 S_1$ & \multicolumn{1}{l}{5811.3}   & \multicolumn{1}{l}{6115.2}   &   6445.2 \\ 
				\multicolumn{1}{l}{}  & \multicolumn{1}{l}{}  &   & \multicolumn{1}{l}{}   & \multicolumn{1}{l}{}   &    \\[-2.45ex]
				\multicolumn{1}{l}{$1^3 L_0$}  & \multicolumn{1}{l}{$1^3 L_0$}  &  $1^5 S_2$ & \multicolumn{1}{l}{6131.8}   & \multicolumn{1}{l}{6467.8}   &   6510.1 \\ \midrule
				\multicolumn{1}{l}{}  & \multicolumn{1}{l}{}  &   & \multicolumn{1}{l}{}   & \multicolumn{1}{l}{}   &    \\[-2.45ex]
				
				\multicolumn{1}{l}{$1^3 L_1$}  & \multicolumn{1}{l}{$1^3 L_1$}  &  $1^1 S_0$ & \multicolumn{1}{l}{5542.9}   & \multicolumn{1}{l}{5760.2}   &  6606.4 \\ 
				\multicolumn{1}{l}{}  & \multicolumn{1}{l}{}  &   & \multicolumn{1}{l}{}   & \multicolumn{1}{l}{}   &    \\[-2.45ex]
				\multicolumn{1}{l}{$1^3 L_1$}  & \multicolumn{1}{l}{$1^3 L_1$}  &  $1^3 S_1$ & \multicolumn{1}{l}{5746.2}   & \multicolumn{1}{l}{5980.6}   &   6643.2 \\ 
				\multicolumn{1}{l}{}  & \multicolumn{1}{l}{}  &   & \multicolumn{1}{l}{}   & \multicolumn{1}{l}{}   &    \\[-2.45ex]
				\multicolumn{1}{l}{$1^3 L_1$}  & \multicolumn{1}{l}{$1^3 L_1$}  &  $1^5 S_2$ & \multicolumn{1}{l}{6074.1}   & \multicolumn{1}{l}{6346.6}   &    6705.6  \\ \midrule
				\multicolumn{1}{l}{}  & \multicolumn{1}{l}{}  &   & \multicolumn{1}{l}{}   & \multicolumn{1}{l}{}   &    \\[-2.45ex]

				\multicolumn{1}{l}{$1^3 L_2$}  & \multicolumn{1}{l}{$1^3 L_2$}  &  $1^1 S_0$ & \multicolumn{1}{l}{5525.4}   & \multicolumn{1}{l}{5725.7}   &  6428.4\\ 
				\multicolumn{1}{l}{}  & \multicolumn{1}{l}{}  &   & \multicolumn{1}{l}{}   & \multicolumn{1}{l}{}   &    \\[-2.45ex]
				\multicolumn{1}{l}{$1^3 L_2$}  & \multicolumn{1}{l}{$1^3 L_2$}  &  $1^3 S_1$ & \multicolumn{1}{l}{5729.8}   & \multicolumn{1}{l}{5948.2}   &   6467.1\\ 
				\multicolumn{1}{l}{}  & \multicolumn{1}{l}{}  &   & \multicolumn{1}{l}{}   & \multicolumn{1}{l}{}   &    \\[-2.45ex]
				\multicolumn{1}{l}{$1^3 L_2$}  & \multicolumn{1}{l}{$1^3 L_2$}  &  $1^5 S_2$ & \multicolumn{1}{l}{6058.9}   & \multicolumn{1}{l}{6316.9}   &    6531.7 \\ \midrule

				\multicolumn{1}{l}{}  & \multicolumn{1}{l}{}  &   & \multicolumn{1}{l}{}   & \multicolumn{1}{l}{}   &    \\[-2.45ex]
				\multicolumn{1}{l}{$1^3 L_0$}  & \multicolumn{1}{l}{$1^3 L_1$}  &  $1^3 S_1$ & \multicolumn{1}{l}{5781.3}   & \multicolumn{1}{l}{6051.0}   & 6507.3\\ 
				\multicolumn{1}{l}{}  & \multicolumn{1}{l}{}  &   & \multicolumn{1}{l}{}   & \multicolumn{1}{l}{}   &    \\[-2.45ex]
				\multicolumn{1}{l}{$1^3 L_0$}  & \multicolumn{1}{l}{$1^3 L_2$}  &  $1^5 S_2$ & \multicolumn{1}{l}{6096.6}   & \multicolumn{1}{l}{6393.9}   &  6456.2\\ 
				\multicolumn{1}{l}{}  & \multicolumn{1}{l}{}  &   & \multicolumn{1}{l}{}   & \multicolumn{1}{l}{}   &    \\[-2.45ex]
				\multicolumn{1}{l}{$1^3 L_1$}  & \multicolumn{1}{l}{$1^3 L_2$}  &  $1^3 S_1$ & \multicolumn{1}{l}{5738.0}   & \multicolumn{1}{l}{5964.4}   & 6555.9\\ 
				\multicolumn{1}{l}{}  & \multicolumn{1}{l}{}  &   & \multicolumn{1}{l}{}   & \multicolumn{1}{l}{}   &    \\[-2.45ex]
				\multicolumn{1}{l}{$1^3 L_1$}  & \multicolumn{1}{l}{$1^3 L_2$}  &  $1^5 S_2$ & \multicolumn{1}{l}{6066.5}   & \multicolumn{1}{l}{6331.7}   & 6618.9 \\ \bottomrule
			\end{tabular}
		\end{center}
	\end{table*}
	\begin{table*}[]\captionof{table}{Possible masses of the all-charm tetraquark for $N=2$.}\label{ResultsTable2}
		\begin{center}
			\begin{tabular}{llllll}
				\toprule
				&                        &   & \multicolumn{3}{l}{}                            \\[-2.4ex] 
				&                        &   & \multicolumn{3}{l}{$M_{12,34}$ (MeV)}                            \\[0.4ex] \cmidrule{4-6}
				\multicolumn{1}{l}{} & \multicolumn{1}{l}{} &  & \multicolumn{1}{l}{} & \multicolumn{1}{l}{} &  \\[-2.4ex]
				\multicolumn{1}{l}{$(N^{2S+1} l_J)_{12}$} & \multicolumn{1}{l}{$(N^{2S+1} l_J)_{34}$} & $(N^{2S+1} l_J)_{12,34}$ & \multicolumn{1}{l}{$MM_{88}$} & \multicolumn{1}{l}{$DA_{6\bar{6}}$} & $DA_{3\bar{3}}$ \\[0.4ex] \midrule			\multicolumn{1}{l}{}  & \multicolumn{1}{l}{}  &   & \multicolumn{1}{l}{}   & \multicolumn{1}{l}{}   &    \\[-2.45ex]
				
				\multicolumn{1}{l}{$1^1 S_0$}  & \multicolumn{1}{l}{$1^1 S_0$}  &  $2^1 S_0$ & \multicolumn{1}{l}{6666.5}   & \multicolumn{1}{l}{6931.5}   &  6558.9  \\ 
				\multicolumn{1}{l}{}  & \multicolumn{1}{l}{}  &   & \multicolumn{1}{l}{}   & \multicolumn{1}{l}{}   &    \\[-2.45ex]
				\multicolumn{1}{l}{$1^1 S_0$}  & \multicolumn{1}{l}{$1^3 S_1$}  & $2^3 S_1$  & \multicolumn{1}{l}{6665.5}   & \multicolumn{1}{l}{6927.8}   &  6583.7\\ 		
				\multicolumn{1}{l}{}  & \multicolumn{1}{l}{}  &   & \multicolumn{1}{l}{}   & \multicolumn{1}{l}{}   &    \\[-2.45ex]
				\multicolumn{1}{l}{$1^3 S_1$}  & \multicolumn{1}{l}{$1^3 S_1$}  & $2^1 S_0$  & \multicolumn{1}{l}{6550.2}   & \multicolumn{1}{l}{6801.5}   &  6573.7 \\ 
				\multicolumn{1}{l}{}  & \multicolumn{1}{l}{}  &   & \multicolumn{1}{l}{}   & \multicolumn{1}{l}{}   &    \\[-2.45ex]
				\multicolumn{1}{l}{$1^3 S_1$}  & \multicolumn{1}{l}{$1^3 S_1$}  & $2^3 S_1$  & \multicolumn{1}{l}{6605.0}   & \multicolumn{1}{l}{6860.6}   &   6591.0 \\ 
				\multicolumn{1}{l}{}  & \multicolumn{1}{l}{}  &   & \multicolumn{1}{l}{}   & \multicolumn{1}{l}{}   &    \\[-2.45ex]
				\multicolumn{1}{l}{$1^3 S_1$}  & \multicolumn{1}{l}{$1^3 S_1$}  & $2^5 S_2$  & \multicolumn{1}{l}{6728.6}   & \multicolumn{1}{l}{6992.9}   &   6626.1 \\ \midrule
				\multicolumn{1}{l}{}  & \multicolumn{1}{l}{}  &   & \multicolumn{1}{l}{}   & \multicolumn{1}{l}{}   &    \\[-2.45ex]

				\multicolumn{1}{l}{$1^1 L_1$}  & \multicolumn{1}{l}{$1^1 L_1$}  &  $2^1 S_0$ & \multicolumn{1}{l}{6793.5}   & \multicolumn{1}{l}{7127.0}   & 7108.4   \\ 
				\multicolumn{1}{l}{}  & \multicolumn{1}{l}{}  &   & \multicolumn{1}{l}{}   & \multicolumn{1}{l}{}   &    \\[-2.45ex]
				\multicolumn{1}{l}{$1^1 L_1$}  & \multicolumn{1}{l}{$1^3 L_0$}  & $2^3 S_1$  & \multicolumn{1}{l}{6823.4}   & \multicolumn{1}{l}{7187.0}   &   7013.5 \\ 	
				\multicolumn{1}{l}{}  & \multicolumn{1}{l}{}  &   & \multicolumn{1}{l}{}   & \multicolumn{1}{l}{}   &    \\[-2.45ex]
				\multicolumn{1}{l}{$1^1 L_1$}  & \multicolumn{1}{l}{$1^3 L_1$}  & $2^3 S_1$  & \multicolumn{1}{l}{6793.4}   & \multicolumn{1}{l}{7124.4}   & 7109.5   \\ 	
				\multicolumn{1}{l}{}  & \multicolumn{1}{l}{}  &   & \multicolumn{1}{l}{}   & \multicolumn{1}{l}{}   &    \\[-2.45ex]
				\multicolumn{1}{l}{$1^1 L_1$}  & \multicolumn{1}{l}{$1^3 L_2$}  & $2^3 S_1$  & \multicolumn{1}{l}{6785.8}   & \multicolumn{1}{l}{7109.2}   &  7110.0 \\ \midrule	
				\multicolumn{1}{l}{}  & \multicolumn{1}{l}{}  &   & \multicolumn{1}{l}{}   & \multicolumn{1}{l}{}   &    \\[-2.45ex]
				
				\multicolumn{1}{l}{$1^3 L_0$}  & \multicolumn{1}{l}{$1^3 L_0$}  &  $2^1 S_0$ & \multicolumn{1}{l}{6744.3}   & \multicolumn{1}{l}{7136.0}   &    6885.9\\ 
				\multicolumn{1}{l}{}  & \multicolumn{1}{l}{}  &   & \multicolumn{1}{l}{}   & \multicolumn{1}{l}{}   &    \\[-2.45ex]
				\multicolumn{1}{l}{$1^3 L_0$}  & \multicolumn{1}{l}{$1^3 L_0$}  &  $2^3 S_1$ & \multicolumn{1}{l}{6795.8}   & \multicolumn{1}{l}{7189.8}   & 6901.8   \\ 
				\multicolumn{1}{l}{}  & \multicolumn{1}{l}{}  &   & \multicolumn{1}{l}{}   & \multicolumn{1}{l}{}   &    \\[-2.45ex]
				\multicolumn{1}{l}{$1^3 L_0$}  & \multicolumn{1}{l}{$1^3 L_0$}  &  $2^5 S_2$ & \multicolumn{1}{l}{6911.2}   & \multicolumn{1}{l}{7308.1}   &  6933.9  \\ \midrule
				\multicolumn{1}{l}{}  & \multicolumn{1}{l}{}  &   & \multicolumn{1}{l}{}   & \multicolumn{1}{l}{}   &    \\[-2.45ex]
				
				\multicolumn{1}{l}{$1^3 L_1$}  & \multicolumn{1}{l}{$1^3 L_1$}  &  $2^1 S_0$ & \multicolumn{1}{l}{6683.8}   & \multicolumn{1}{l}{7008.6}   &    7080.5\\ 
				\multicolumn{1}{l}{}  & \multicolumn{1}{l}{}  &   & \multicolumn{1}{l}{}   & \multicolumn{1}{l}{}   &    \\[-2.45ex]
				\multicolumn{1}{l}{$1^3 L_1$}  & \multicolumn{1}{l}{$1^3 L_1$}  &  $2^3 S_1$ & \multicolumn{1}{l}{6736.5}   & \multicolumn{1}{l}{7064.6}   & 7095.4  \\ 
				\multicolumn{1}{l}{}  & \multicolumn{1}{l}{}  &   & \multicolumn{1}{l}{}   & \multicolumn{1}{l}{}   &    \\[-2.45ex]
				\multicolumn{1}{l}{$1^3 L_1$}  & \multicolumn{1}{l}{$1^3 L_1$}  &  $2^5 S_2$ & \multicolumn{1}{l}{6854.8}   & \multicolumn{1}{l}{7188.4}   &  7125.8  \\ \midrule
				\multicolumn{1}{l}{}  & \multicolumn{1}{l}{}  &   & \multicolumn{1}{l}{}   & \multicolumn{1}{l}{}   &    \\[-2.45ex]
				
				\multicolumn{1}{l}{$1^3 L_2$}  & \multicolumn{1}{l}{$1^3 L_2$}  &  $2^1 S_0$ & \multicolumn{1}{l}{6668.0}   & \multicolumn{1}{l}{6977.4}   &    6907.5\\ 
				\multicolumn{1}{l}{}  & \multicolumn{1}{l}{}  &   & \multicolumn{1}{l}{}   & \multicolumn{1}{l}{}   &    \\[-2.45ex]
				\multicolumn{1}{l}{$1^3 L_2$}  & \multicolumn{1}{l}{$1^3 L_2$}  &  $2^3 S_1$ & \multicolumn{1}{l}{6720.9}   & \multicolumn{1}{l}{7033.8}   & 6923.2 \\ 
				\multicolumn{1}{l}{}  & \multicolumn{1}{l}{}  &   & \multicolumn{1}{l}{}   & \multicolumn{1}{l}{}   &    \\[-2.45ex]
				\multicolumn{1}{l}{$1^3 L_2$}  & \multicolumn{1}{l}{$1^3 L_2$}  &  $2^5 S_2$ & \multicolumn{1}{l}{6839.7}   & \multicolumn{1}{l}{7158.8}   &  6955.2 \\ \midrule
				\multicolumn{1}{l}{}  & \multicolumn{1}{l}{}  &   & \multicolumn{1}{l}{}   & \multicolumn{1}{l}{}   &    \\[-2.45ex]
				
				\multicolumn{1}{l}{$1^3 L_0$}  & \multicolumn{1}{l}{$1^3 L_1$}  &  $2^3 S_1$ & \multicolumn{1}{l}{6767.3}   & \multicolumn{1}{l}{7128.6}   &  6983.7 \\ 
				\multicolumn{1}{l}{}  & \multicolumn{1}{l}{}  &   & \multicolumn{1}{l}{}   & \multicolumn{1}{l}{}   &    \\[-2.45ex]
				\multicolumn{1}{l}{$1^3 L_0$}  & \multicolumn{1}{l}{$1^3 L_2$}  &  $2^5 S_2$ & \multicolumn{1}{l}{6876.3}   & \multicolumn{1}{l}{7234.5}   &  6912.5 \\ 
				\multicolumn{1}{l}{}  & \multicolumn{1}{l}{}  &   & \multicolumn{1}{l}{}   & \multicolumn{1}{l}{}   &    \\[-2.45ex]
				\multicolumn{1}{l}{$1^3 L_1$}  & \multicolumn{1}{l}{$1^3 L_2$}  &  $2^3 S_1$ & \multicolumn{1}{l}{6728.7}   & \multicolumn{1}{l}{7049.2}   &  7009.8\\ 
				\multicolumn{1}{l}{}  & \multicolumn{1}{l}{}  &   & \multicolumn{1}{l}{}   & \multicolumn{1}{l}{}   &    \\[-2.45ex]
				\multicolumn{1}{l}{$1^3 L_1$}  & \multicolumn{1}{l}{$1^3 L_2$}  &  $2^5 S_2$ & \multicolumn{1}{l}{6847.3}   & \multicolumn{1}{l}{7173.6}   & 7040.8 \\  \bottomrule
			\end{tabular}
		\end{center}
	\end{table*}

\end{document}